# Surface superconductivity in the type II Weyl semimetal $TaIrTe_4$

Ying Xing[1,2†], Zhibin Shao[3,†], Jun Ge[2], Jiawei Luo[2], Jinhua Wang[4,5], Zengwei Zhu[4,5], Jun Liu[6], Yong Wang[6], Zhiying Zhao[7,11], Jiaqiang Yan[7,8], David Mandrus[7,8], Binghai Yan[9], Xiong-Jun Liu[2,10,12,13,*], Minghu Pan[3,4,*], Jian Wang[2,10,12,13,*]

[1] *Department of Materials Science and Engineering, School of New Energy and Materials, China University of Petroleum, Beijing 102249, China;*

[2] *International Center for Quantum Materials, School of Physics, Peking University, Beijing 100871, China;*

[3] *School of Physics and Information Technology, Shaanxi Normal University, Xi'an 710119, China;*

[4] *School of Physics, Huazhong University of Science and Technology, Wuhan 430074, China;*

[5] *Wuhan National High Magnetic Field Center, Huazhong University of Science and Technology, Wuhan 430074, China;*

[6] *Center of Electron Microscopy, State Key Laboratory of Silicon Materials, School of Materials Science and Engineering, Zhejiang University, Hangzhou, 310027, China;*

[7] *Department of Materials Science and Engineering, University of Tennessee, Knoxville, Tennessee 37996, USA;*

[8] *Materials Science and Technology Division, Oak Ridge National Laboratory, Oak Ridge, Tennessee 37831, USA;*

[9] *Department of Condensed Matter Physics, Weizmann Institute of Science, Rehovot, 7610001, Israel;*

[10] *CAS Center for Excellence in Topological Quantum Computation, University of Chinese Academy of Sciences, Beijing 100190, China;*

[11] *Department of Physics and Astronomy, University of Tennessee, Knoxville, TN 37996, USA;*

[12] *Beijing Academy of Quantum Information Sciences, Beijing 100193, China;*

[13] *Collaborative Innovation Center of Quantum Matter, Beijing 100871, China*

†These authors contributed equally to this work.

*e-mail: jianwangphysics@pku.edu.cn (J.W.); minghupan@hust.edu.cn (M.P.); xiongjunliu@pku.edu.cn(X.L.)

## ABSTRACT

The search for unconventional superconductivity in Weyl semimetal materials is currently an exciting pursuit, since such superconducting phases could potentially be topologically nontrivial and host exotic Majorana modes. The layered material $TaIrTe_4$ is a newly predicted time-reversal invariant type II Weyl semimetal with minimum number of Weyl points. Here, we report the discovery of surface superconductivity in Weyl semimetal $TaIrTe_4$. Our scanning tunneling microscopy/spectroscopy (STM/S) visualizes Fermi arc surface states of $TaIrTe_4$ that are consistent with the previous angle-resolved photoemission spectroscopy (ARPES) results. By a systematic study based on STS at ultralow temperature, we observe uniform superconducting gaps on the sample surface. The superconductivity is further confirmed by electrical transport measurements at ultralow temperature, with an onset transition temperature ($T_c$) up to 1.54 K being observed. The normalized upper critical field $h^*(T/T_c)$ behavior and the stability of the superconductivity against the ferromagnet indicate that the discovered superconductivity is unconventional with the *p*-wave pairing. The systematic STS, thickness and angular dependent transport measurements reveal that the detected superconductivity is quasi-one-dimensional (quasi-1D) and occurs in the surface states. The discovery of the surface superconductivity in $TaIrTe_4$ provides a new novel platform to explore topological superconductivity and Majorana modes.

## INTRODUCTION

Weyl semimetals, which possess nodal points in the bulk and Fermi arc states in the surface, have generated considerable research interest in the recent years [1-12]. The chirality of Weyl fermions is responsible for a few novel transport phenomena, such as the chiral anomaly. On the surface of Weyl semimetals, universal signatures of topological Fermi arcs in quasi-particle interference were theoretically predicted [13] and experimentally observed by STM [14-16]. Moreover, the theoretical studies have shown that the presence of superconductivity in Weyl semimetal may lead to a bunch of novel topological phases, including the time-reversal invariant topological superconductor [17], Fulde-Ferrell-Larkin-Ovchinnikov superconductors [18-20], and chiral non-Abelian Majorana fermions protected by second Chern numbers [21]. These predictions suggest that turning a Weyl semimetal into superconducting state may provide a promising way to explore topological superconductivity and Majorana modes, which can be applied to topological quantum computation [22-24].

Experimentally, superconductivity has been observed in both type I and type II Weyl semimetals, such as tip induced superconductivity on TaAs [25], pressure induced superconductivity on TaP [26], as well as $T_d$ phase $WTe_2$ (pressure driven) [27,28] and $MoTe_2$ crystals (without pressure) [16,29]. However, in these Weyl semimetals, the number of Weyl points is 8 or even 24, more than the minimal number of Weyl points allowed for a time-reversal invariant Weyl semimetal, which leads to complicated band structures and hinders further studies. Therefore, to observe superconductivity in simpler Weyl semimetals possessing minimal number of Weyl points is highly desired.

Following the first-principle calculations by K. Koeperni *et al.* [30], $TaIrTe_4$ hosts only four well-separated Weyl points, which is the minimal number in a Weyl semimetal with time-reversal symmetry. The Fermi arcs connecting Weyl nodes of opposite chirality in $TaIrTe_4$ extend to about ⅓ of the surface Brillouin zone in the *b* direction. This large momentum-space separation makes $TaIrTe_4$ quite favorable for exploring the Fermi arcs spectroscopically and the important transport properties. The exotic surface states supporting the quasi-1D Fermi arcs have been observed by ARPES[31]. Fermi arcs as well as Weyl nodes in the bulk of $TaIrTe_4$ have been identified directly by pump-probe ARPES[32]. The Weyl points and Fermi arcs are found to live at 50 ~ 100 meV above Fermi energy. If the noncentrosymmetric Weyl material $TaIrTe_4$ can be superconducting, it would stimulate further investigations on the superconductivity in topological materials and long-sought-after topological superconductors.

In this work, we perform STM and STS and electrical transport studies of the ternary compound $TaIrTe_4$ single crystal down to 0.06 K with a high magnetic field up to 54.5 T. The Fermi arc surface states and superconducting gap are discovered by STM and STS studies at ultralow temperatures. The detected unconventional superconductivity is further verified to only occur on the surface of $TaIrTe_4$ by electrical transport measurements. The observed unconventional surface superconductivity is found to exhibit quasi-1D and topologically non-trivial characteristics, which demonstrate that $TaIrTe_4$ is a unique candidate for topological superconductors.

## RESULTS

### Sample characterization of $TaIrTe_4$ crystals

Single crystals of $TaIrTe_4$ were synthesized from excess Te flux. The crystal is needle-like morphology and grows preferentially along the [100] direction (the length direction). The width direction is along the [010] and the cleavage surface of the crystal is the (001) plane. The high structural quality of the sample was confirmed by X-ray diffraction (XRD), high-resolution scanning transmission electron microscopy (HRSTEM) and STM. Fig. 1a shows the XRD from a $TaIrTe_4$ crystal oriented with the scattering vector perpendicular to the (001) plane. The inset is the morphology of a representative crystal looking down from the [001] direction. The atomic HRSTEM image (Fig. 1b) and selected area electron diffraction pattern (Fig. S1) further confirm the crystalline property of our $TaIrTe_4$ crystals. The obtained lattice parameters $\boldsymbol{a}$ = 0.375 nm, $\boldsymbol{b}$ = 1.246 nm, $\boldsymbol{c}$ = 1.304 nm, which agree with the previous report on $TaIrTe_4$[33]. STM investigation shows uniform large-scaled periodical 1D stripes on a cleaved $TaIrTe_4$ surface. Several bright spots appear on the flat Te terrace, which can be attributed to some adatoms from the upper Te plane left on the terrace during cleaving. From a zoom-in image (Fig. 1d), a unidirectional stripe was observed. The fast Fourier transform (FFT) in Fig. 1f used to calibrate the modulation of 1D pattern, reveals the periodicity of stripe is about 1.2 nm in real space. And this is in well agreement with lattice parameter of the $\boldsymbol{b}$ direction (12.421Å), which suggests no reconstruction in $\boldsymbol{b}$ direction.

### Fermi arc surface states and superconducting gap detected by STM/STS

Quasiparticle interference (QPI) based on spectroscopic-imaging STM, has shown success in identifying the topological surface states of topological insulator[34,35] and topological Fermi arc states of Weyl semimetals TaAs[36], $MoTe_2$[14] , $MoTe_{2-x}S_x$[16] and $Mo_{0.66}W_{0.34}Te_2$[37]. In the surface Brillouin zone, the extremal pairs of $\boldsymbol{k}_i$ and $\boldsymbol{k}_f$ on a two-dimensional constant energy contour, where $\boldsymbol{k}_i$ and $\boldsymbol{k}_f$ are the initial and final wavevectors, contribute dominantly to the spatial interference pattern of the local density of states. The features in Fourier transform of d$I$/d$V$ mapping correspond to the scattering vector $\boldsymbol{q} = \boldsymbol{k}_i - \boldsymbol{k}_f$ of the extremal pairs.

To detect Fermi arc states in our $TaIrTe_4$ single crystals, STS mappings were performed at the (001) surface of $TaIrTe_4$ single crystals. Fig. S2b-h show d$I$/d$V$ mappings taken with various biases from 20 meV to 80 meV. Fig. 2a-g show the fast Fourier transform (FFT) of the d$I$/d$V$ maps between 20 meV and 80 meV. Four arcs located inside the first Brillouin zone were revealed by the QPI imaging at the energy of 80 meV (Fig. 2g). For a pair of topological Fermi arcs, three scattering wavevectors (Fig. 2h), labelled $q_1$, $q_2$ and $q_3$, might be expected to appear in QPI. Among them, $q_2$ is forbidden due to the requirement of the time-reversal symmetry in the system[30]. $q_3$ does not correspond to the observed features discussed above because this vector has very small length in $k$ space and stays very close to the center. The scattering wavevectors should generate visible features of four arcs. Such features are clearly resolved at 80 mV in our experiments as indicated by yellow arrows in Fig. 2g, and become obscure at lower energies (50-70 meV), eventually vanish below 40 meV. Previously theoretical study has reported that Fermi arc locates at a narrow energy range between 50 and 82.7 meV in $TaIrTe_4$[30]. As the energy location moves upward from 50 meV, Fermi arc partially separates out from bulk bands and completely appears at the energy of Wely nodes (82.7 meV). The mixing of Fermi arc and bulk band at lower energies will lead to the obscureness of Fermi arc imaging, which is consistent with our experimental observation. This is a direct and strong experimental evidence for the existence of the topological surface states.

Fig. 3a gives the typical STM topographic image of the cleaved surface obtained at the bias of -20 mV and at temperature of 4 K. Periodical atomic chains along $a$ direction are clearly observed on the surface, confirming the quasi-1D characteristic of $TaIrTe_4$. Compared with 4 K, the crystal structure at 0.4 K remains undistorted, which excludes the possibility of structure phase transition occurring at low temperatures. The d$I$/d$V$ spectrum taken at 0.4 K displays a clear signature of superconducting gap with two conductance peaks at gap edges, as shown in Fig. 3b. The spectrum exhibits a superconducting gap ($\Delta$) of 2.1 meV defined by half the distance between the two conductance peaks. The superconducting gap is uniform on the whole cleaved surface (see Fig. S3 and S4). After macroscopically changing the locations of STM scanning, similar topographic images and superconductivity were observed reproducibly. Fig. 3c shows the temperature evolution of d$I$/d$V$ spectra measured from 0.4 K to 1.28 K. As the temperature increases, the dip at zero bias is reduced and the gap almost vanishes near 1.28 K. The ratio $\Delta(0)/k_B T_c$ ($k_B$ is the Boltzmann constant) is estimated about 19.05, which is much larger than that of weak coupling BCS superconductors and reminiscent of the possibility of topological superconductivity[16,38,39]. Magnetic field dependence of d$I$/d$V$ spectra are shown in Fig. 3d. The superconducting gap decreases with the increasing field and almost vanishes near 0.25 T, exhibiting the typical feature of superconductivity. Both critical values (1.28 K and 0.25 T) match well with the results of our transport measurement (Fig. 4). Furthermore, we locate a terrace edge that is perpendicular with the direction of 1D atomic rows and perform a line spectroscopic survey along a 1D atomic row by crossing the broken end (blue dashed line in lower panel of Fig. 3f). The d$I$/d$V$ spectra along the 1D Ta-Ir chain show that superconducting gap becomes smaller and shallower by approaching to the broken end and finally almost vanishes (Fig. 3e). It is worth mentioning that superconductivity can be observed on every terrace of the sample. The crucial dependence of the superconductivity on defects in the Ta-Ir chains again suggests that the pairing order might be unconventional, in contrast to the conventional *s*-wave order which is stable against to defects.

## Electrical transport evidences of Quasi−1D superconductivity and quantum oscillations

To further demonstrate and explore the observed superconductivity in $TaIrTe_4$ single crystals, transport measurements at ultralow temperature were performed at the (001) surface of $TaIrTe_4$ single crystals. The $TaIrTe_4$ single crystal samples were cleaved to a smooth and fresh surface for transport measurements. More than 10 samples are studied and all samples exhibit consistent results. Fig. 4a shows the resistivity of sample 1 (S1) as a function of temperature ($T$) from 2 K to 300 K. The resistivity exhibits metallic-like behavior and tends to saturate at 10 K with a residual resistivity ratio (RRR) of 6.8 (the resistivity at room temperature over the resistivity at 2 K). Interestingly, when upon further cooling, an evident resistivity drop appears at about 1.54 K (Fig. 4b). When applying perpendicular magnetic field ($B$//***c*** axis), the resistivity drop shifts to lower temperatures as the field increases and completely suppressed at around 0.4 T. This is a typical superconducting behavior although no zero resistance is observed down to 0.06 K and the proportion of resistivity drop is ~ 44% (Fig. 4b). Magnetotransport measurements for $B$//***c*** axis (Fig. 4c), $B$//***b*** axis (Fig. 4d), and $B$//***a*** axis (Fig. 4e) were carried out at various temperatures from 0.08 to 2.0 K. It is evident that superconductivity at $B$//***a*** axis varies differently from the other two directions. For example, $B_{c2}$ is around 0.5 T at 0.1 K for both $B$//***c*** axis and $B$//***b*** axis, substantially smaller than $B_c > 1.5$ T for the $B$//***a*** axis situation (Fig. 4f). This agrees well with the observed $B_{c2}$ (0.25 T at 0.4 K for $B$//***c*** axis) from STS measurements (see Fig. 3d). Since zigzag Ta-Ir chains along ***a*** direction of $TaIrTe_4$, the difference of $B_{c2}$ may originate primarily from the anisotropy of the sample, which causes quasi-1D superconductivity[40, 41]. Besides the observed anisotropic superconductivity, when the temperature is above $T_c$, the pronounced anisotropic MR at 2 K up to 15 T is also detected, which further confirms the anisotropic characteristic of $TaIrTe_4$, as shown in Fig. S5. For quasi-1D superconductors, below $T_c$ phase slips can give rise to broad superconducting transition with the residual resistance[42], which may explain our observations. Another possible scenario is that the superconductivity occurs in the surface states which are helical states for a time-reversal invariant Weyl semimetal with dispersions along ***a***-axis, leading to the quasi-1D *p*-wave superconducting phases. Similar resistivity drops are observed in two other $TaIrTe_4$ samples, with onset $T_c$ (where resistivity starts to drop) from 1.19 K to 1.38 K (Fig. S6, consistent with the $T_c$ ~1.28 K obtained from STS result), confirming the observed superconductivity in $TaIrTe_4$ crystals. Further measurements show that the $T_c$ of different regions in the same $TaIrTe_4$ sample exhibits consistent superconducting behavior, which excludes the macroscopic superconducting phase separation in $TaIrTe_4$ crystals (Fig. S7). The reduced critical field $h^*$ equals $B_{c2}/[T_c(-dB_{c2}/dT|_{T=Tc})]$, which is calculated to compare with known models for *s*-wave superconductors (Werthamer–Helfand–Hohenberg theory, WHH, $h^*(0) \approx 0.7$[43]) and spin-triplet *p*-wave superconductors ($h^*(0) \approx 0.8$[44]). $B_{c2}$ is defined as the field above which the $TaIrTe_4$ sample becomes the normal state. Obviously, the $h^*(T/T_c)$ relation is close to that of a polar *p*-wave state, suggesting the possibility of unconventional superconducting paring symmetry in $TaIrTe_4$ as shown in the inset of Fig. 4c. Critical current ($I_c$) is another key feature of superconductors. Fig. S8a and b depict $R(I)$ characteristics of S3 (30 μm thick) at different temperatures and magnetic fields. At 0.3 K and 0 T, as the current increasing, the sample is gradually tuned from superconducting state to normal state. The $I_c$ is suppressed by both temperature and magnetic field, which provides another evidence of superconductivity in $TaIrTe_4$. In addition, the existence of *p*-wave superconductivity is backed by the essential symmetry consideration, in which both the bulk and surface of the studied material break the inversion symmetry and thus, allow the spin-triplet pairing[45].

We also investigate the MR at ultrahigh pulsed magnetic field. Fig. 4g presents the magneto-resistivity of sample 4 (S4) up to 54.5 T at various temperatures from 0.3 K to 20 K. At 0.3 K, superconducting drop below 0.5 T coexisting with pronounced SdH oscillations is observed. By subtracting a polynomial background, the oscillatory components $\Delta\rho$ vs. $1/B$ at different temperatures are plotted in Fig. 4h, with the $\Delta\rho$ oscillations periodic in $1/B$. In Fig. 4i, the oscillatory components $\Delta\rho$ are analyzed by employing FFT at various temperatures from 0.3 K to 20 K. The FFT spectrums exhibit two oscillating frequencies at 64.1 T, 104.5 T. The second harmonics 209.7 T, approximately 2 times of 104.4 T, is likely due to spin splitting. We use two independent parameters to fit the effective mass $m^*$. The $m^*$~$0.349m_e$ and $m^*$~$0.412m_e$ are obtained for 64.1 T and 104.5 T from temperature dependence of the oscillation amplitudes fitted by Lifshitz Kosevich formula. The inset of Fig. 4i is the plot of Landau index $n$ vs. $1/B$, from which the quantum limit field is estimated to 95.32 T. The maxima of the $\Delta\rho$ are assigned to be the integer indices (solid circles) and the minima of $\Delta\rho$ are plotted by open circles in the diagram as half-integer indices. A linear extrapolation of $n$ versus $1/B$ plot gives the intercept value close to -0.11. The observed results are further confirmed in sample 5 (S5) (Fig. S9). The evolution of the Fermi surface in different directions can be revealed by the angular-dependent magnetic quantum oscillations (see Fig. S10). The angle-resolved SdH FFT peaks from $B_{[010]}$ to $B_{[001]}$ axis give the complexity and anisotropy of Fermi surface, similar to the previous report by torque measurements[46].

## Evidences for surface superconductivity

The low proportion of resistivity drop in $TaIrTe_4$ indicates the tiny superconducting volume fraction, and does not support the bulk superconductivity. Our STS results show uniform superconducting gap on the surface of $TaIrTe_4$. More importantly, it is found that the $I_c$ remains of same magnitude when reducing the thickness of sample 3 from 30 μm to 6 μm (see Fig. 5a and Fig. S8). In thickness control experiments, both the sample width (120 μm) and length (the distance between two voltage electrodes: 560 μm for 6 μm thick sample and 590 μm for 30 μm thick sample) are nearly consistent. This observation provides a strong evidence of the surface superconductivity, rather than the bulk superconductivity, since the $I_c$ of a bulk superconductor decreases proportionately as the thickness decreases. Note that the surface of $TaIrTe_4$ is sensitive to atmosphere environment and the degree of surface oxidation is not exactly the same for different samples when carrying out transport measurements. Thus, the 25% discrepancy in $I_c$ should be from the different degradation degrees of sample surface as the surface of 6 μm thick sample is fresher than 30 μm thick sample. Fig. 5b shows the angular dependence of the upper critical field $B_{c2}$ ($\theta$ represents the angle between the ***a*** axis and applied magnetic field directions in ***ac*** plane) at 0.5 K. A cusp-like peak is clearly resolved at $\theta = 90°$ (***B***//***a***) and is qualitatively distinct from the 3D mass model but can be described by the 2D Tinkham model. Such behavior further suggests that the superconductivity in $TaIrTe_4$ is from the surface. These results fully exclude the scenario of the bulk superconductivity and support the surface superconductivity explanation. We notice that $\gamma = H_{C2}^{//}/H_{C2}^{\perp}$ value is smaller compared with typical 2D superconductors[47,48], which could be attributed to the quasi-1D modulation on the surface superconductivity[41].

The suppression of the superconductivity close to the terrace edge as shown in Fig. 3f is consistent with the surface superconductivity of the Fermi arc states. The terrace edge serves as the strong dislocation on the sample surface. Nearby the terrace edge the Fermi arc surface states are pushed into the deeper layers from the outmost layer, leading to a small or vanishing local density of states on the edge. For this reason, the superconductivity can be naturally suppressed. Due to the helical behavior of the Fermi arc states, the surface superconductivity is potentially *p*-wave type and topologically nontrivial. This *p*-wave feature can be further confirmed by depositing a 10 nm thick ferromagnetic Ni film on the surface of bulk $TaIrTe_4$. The MR hysteresis of $TaIrTe_4$ with Ni film indicates the magnetic property of the deposited Ni film (Fig. S11). Interestingly, the magnetic Ni film has little effect on the onset $T_c$ of $TaIrTe_4$ (Fig. 5d), which supports the *p*-wave like or topological superconductivity from the surface state together with the fitting for critical field vs. temperature behavior, suppression of the superconductivity at the terrace edge, and the detected Fermi arc surface state.

## CONCLUSION

In conclusion, we have observed the novel superconductivity in type II Weyl semimetal $TaIrTe_4$ by both low temperature STM/STS and transport studies. The uniform superconducting gap on the sample surface, residual resistance below $T_c$, nearly thickness-independent ultralow critical current, and anisotropic upper critical field behavior indicate that the superconductivity occurs in the surface states. Moreover, the edge-sensitive superconducting gap, the critical field vs. temperature behavior, the topological Fermi arc surface states, and the stability of the superconductivity against the magnetization support the *p*-wave-like topological nature of the quasi-1D superconductivity. Our results suggest that $TaIrTe_4$ is a new promising candidate of topological superconductors.

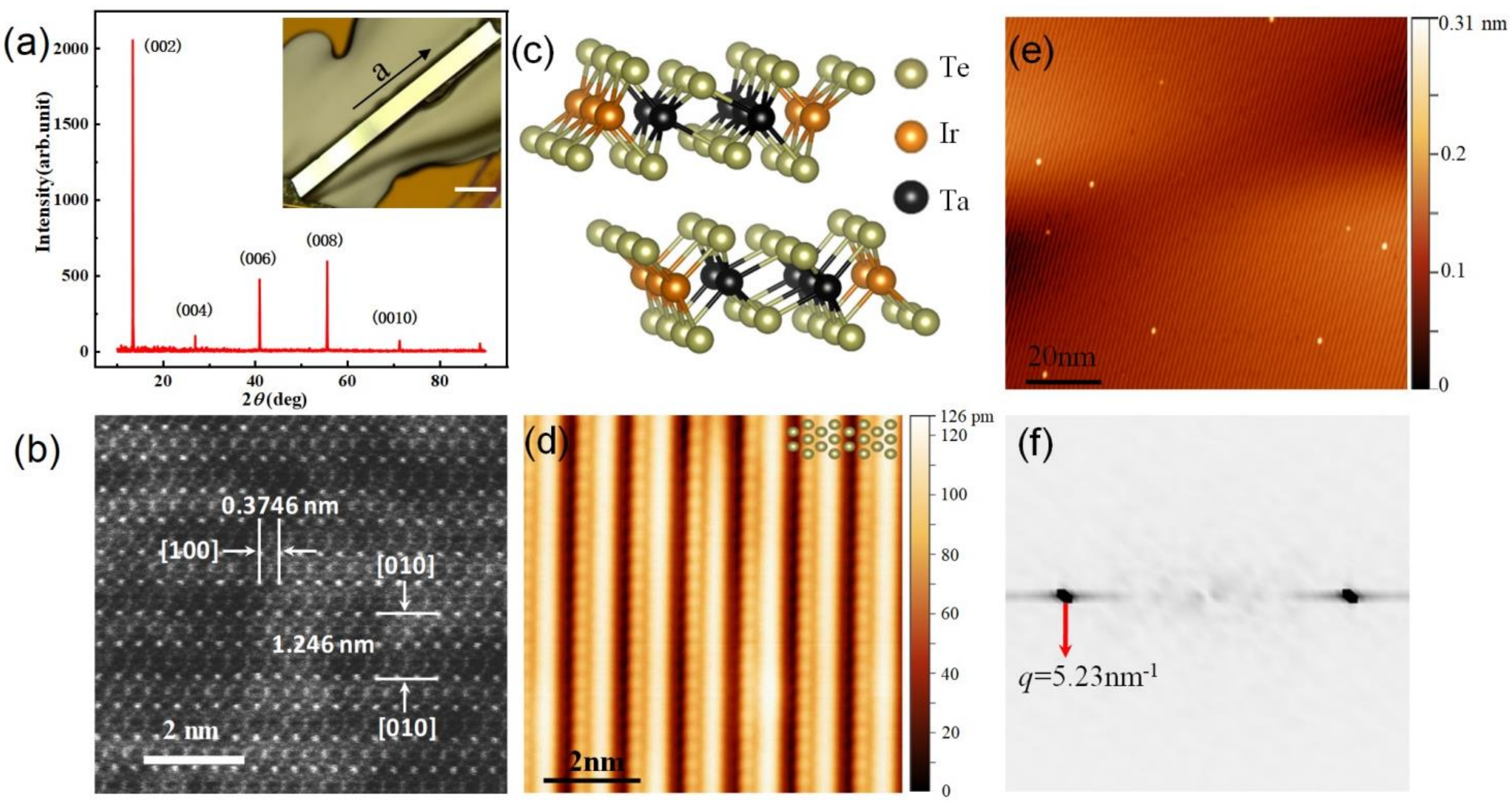

**Figure 1**. Characterization of Weyl semimetal $TaIrTe_4$. (a) The XRD pattern from the basal surface of $TaIrTe_4$ only shows $(002)_n$ reflections, which indicates the measured crystal plane of the crystal is (001) plane. Inset: optical image of a typical $TaIrTe_4$ single crystal. The scale bar is 200 μm. (b) HRSTEM image of the $TaIrTe_4$, showing atomic structure. (c) Schematics of crystal structure of $TaIrTe_4$. (d) and (e) STM images of the fresh cleaved surface of $TaIrTe_4$ with the setting parameters of $V_{bias}$ = 25 mV, $I_{set}$ = 300 pA and $V_{bias}$ = 2.5 V, $I_{set}$ = 20 pA, respectively. (f)Fast Fourier transform image of (d).

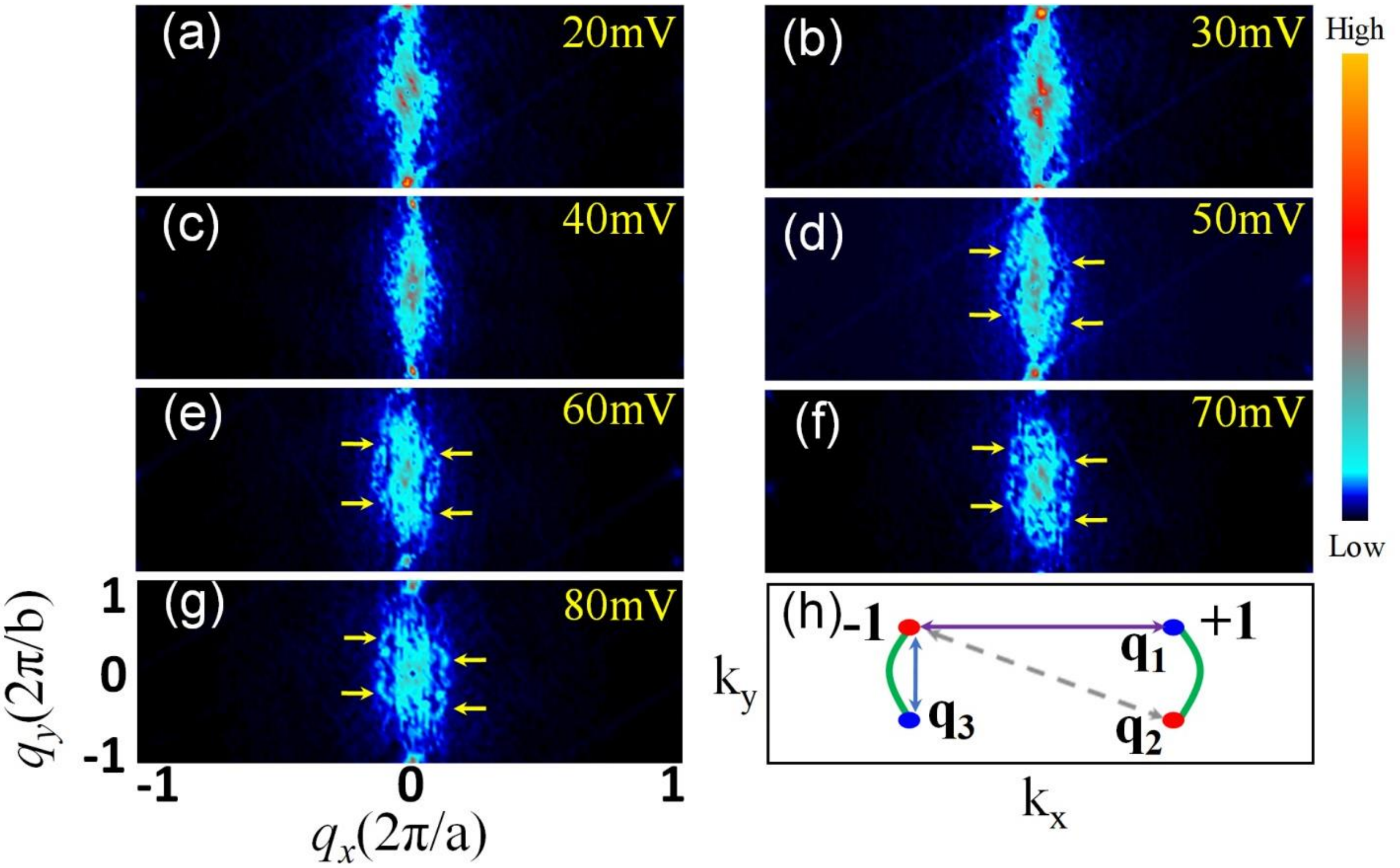

**Figure 2**. Fermi arc states of $TaIrTe_4$ detected by STS. (a-g) Fourier transform of d*I*/d*V* maps at indicated energies. All maps were taken with set point of 250 pA. The resolution is 512×512 pixels. Yellow arrows indicate the interference pattern due to topological surface states. (h) Scattering geometry of Fermi arcs in *k* space.

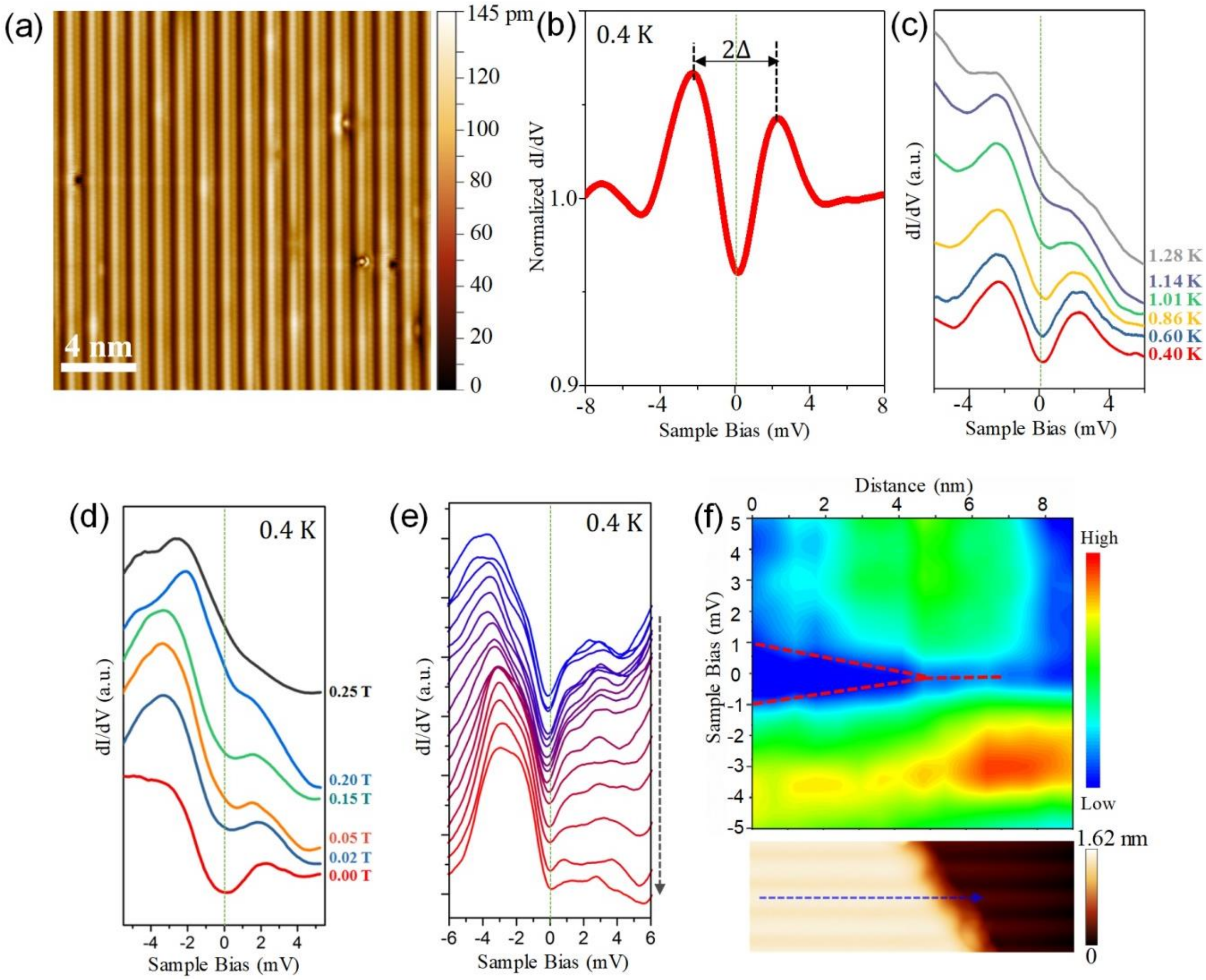

**Figure 3**. The superconductivity in $TaIrTe_4$ detected by STM/STS. (a) The typical STM topographic image of the cleaved surface of $TaIrTe_4$ (temperature: 4 K, bias voltage: -50 mV, tunneling current: 300 pA, 20 × 20 nm$^2$), showing quasi-1D structure. (b) The normalized differential conductance d*I*/d*V* spectrum measured on the terrace of $TaIrTe_4$ surface at 0.4 K, showing a superconducting gap, with the value of 2.1 meV. (c) Temperature dependence of d*I*/d*V* spectra from 0.4 K to 1.28 K. Spectra measured at different temperatures are shifted vertically for clarity. (d) Magnetic field dependence of d*I*/d*V* spectra from 0 T to 0.25 T at 0.4 K. (e) From top to bottom: the d*I*/d*V* spectra acquired along 1D atomic row at 0.4 K by crossing a terrace edge showing in lower panel of (f). (f) A colorful plot of the spectroscopic survey measured along blue dashed line showed in lower panel. Lower: the STM image shows the 1D atomic row with a broken end. All d*I*/d*V* tunneling spectra are measured with a bias voltage of -10 mV, a tunneling current of 500 pA. The bias modulation is set at 150 μV.

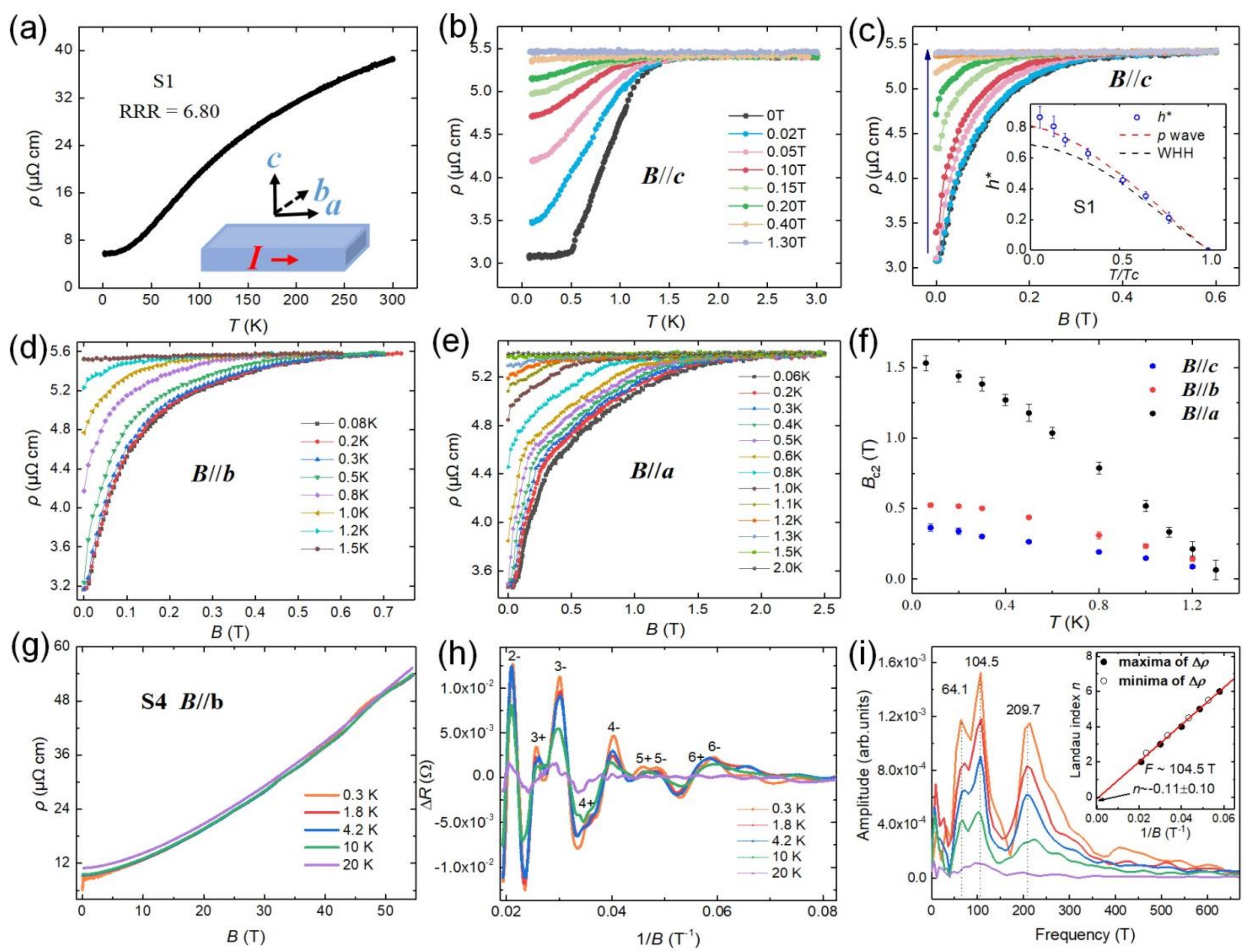

**Figure 4**. Electric transport properties of $TaIrTe_4$ single crystal (S1) showing quasi-1D superconductivity. (a) Resistivity as a function of temperature between 2 K and 300 K. Inset shows schematic structure of crystal orientation in $TaIrTe_4$ samples. (b) $\rho(T)$ curves at different perpendicular magnetic fields ($B//\boldsymbol{c}$ axis) from 0 T to 1.30 T. At 0 T, the sample resistivity begins to drop at 1.54 K ($T_c$). (c) $\rho(B//\boldsymbol{c})$ curves at various temperatures at 0.08 K, 0.2 K, 0.3 K, 0.5 K, 0.8 K, 1.0 K, 1.2 K, 1.5 K and 2.0 K. The arrow indicates the increasing temperature. Inset: normalized upper critical field $h^* = B_{c2}/[T_c(-dB_{c2}/dT|_{T=Tc})]$ as a function of normalized temperature $T/T_c$. The red dashed line indicates the expectation for a polar *p*-wave state. The black dashed line indicates the WHH theory for *s*-wave superconductor. (d) $\rho(B//\boldsymbol{b})$ curves and (e) $\rho(B//\boldsymbol{a})$ curves at various temperatures. (f) Onset critical magnetic fields $B_{c2}$ for $B//\boldsymbol{b}$, $B//\boldsymbol{a}$ and $B//\boldsymbol{c}$ as a function of temperature $T$. (g) Quantum oscillations in $TaIrTe_4$ single crystals (S4) at $B_{[010]}$ direction. Magnetic field (up to 54.5 T) dependence of resistivity at different temperatures with magnetic field perpendicular to the *ac* plane ($B//\boldsymbol{b}$). (h) The oscillatory component of $\Delta\rho$ extracted from $\rho$ by subtracting a polynomial background, as a function of $1/B$ at various temperatures. (i) FFT analysis with two major frequencies (64.1 T and 104.5 T) for $\Delta\rho$ vs. $1/B$ in (h). Inset: Landau index ($n$) as a function of $1/B$. $B_{lim}$ is estimated to 95.32 T.

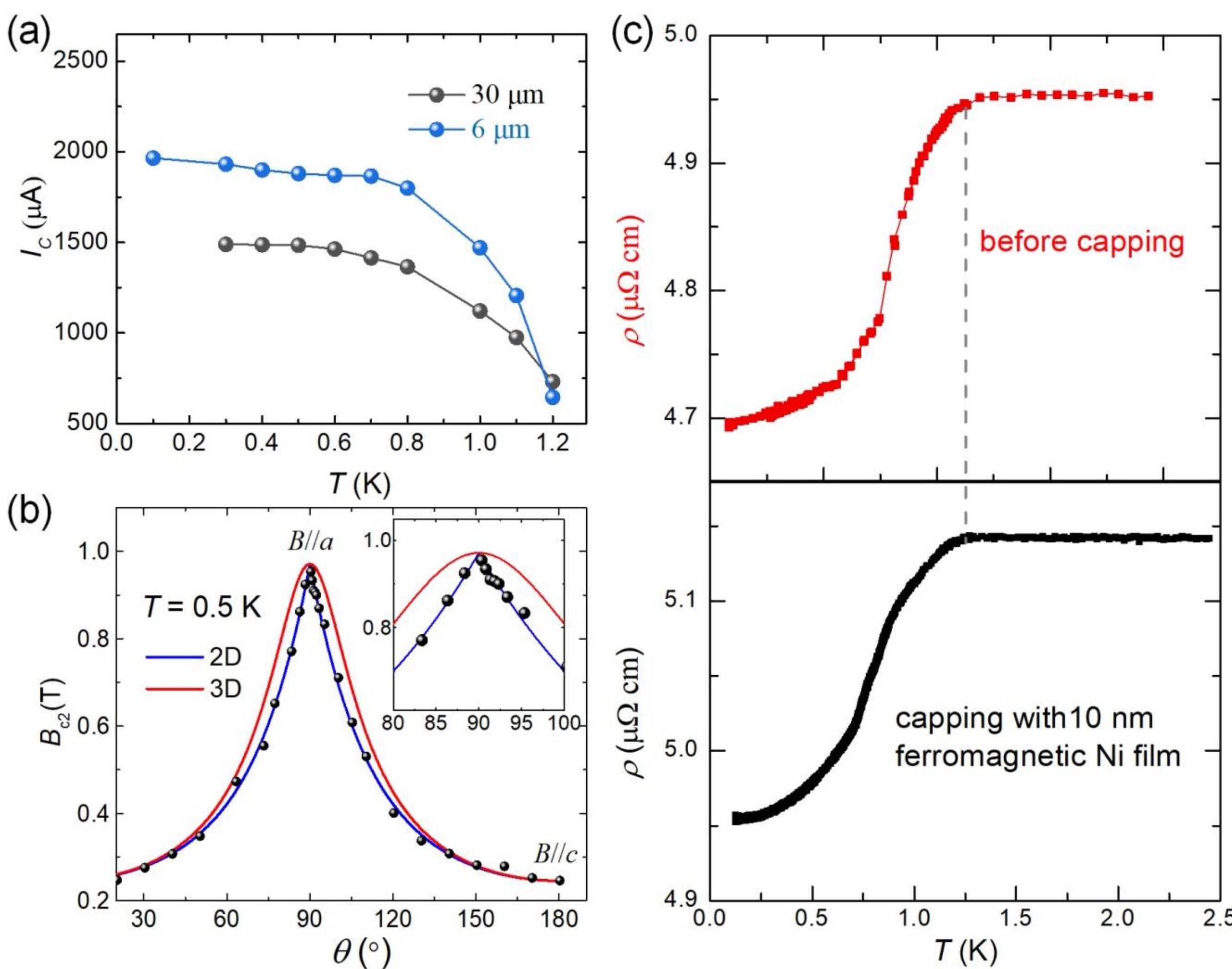

**Figure 5**. Evidences for unconventional surface superconductivity on $TaIrTe_4$. (a) Temperature dependence of $I_c$ of two $TaIrTe_4$ samples with two different thickness (30 μm, 6 μm). These samples were obtained by mechanical exfoliation from one sample (Sample 3). (b) Angular dependence of the upper critical field $B_{c2}$ at 0.5 K. The blue and red line are the theoretical representations of $B_{c2}(\theta)$ using the 2D Tinkham formula $\left(H_{C2}(\theta)\sin\theta/H_{C2}^{//}\right)^2+\left|H_{C2}(\theta)\cos\theta/H_{C2}^{\perp}\right|=1$ and the 3D anisotropic mass model $H_{C2}(\theta)=H_{C2}^{//}/\left(\sin^2\theta+\gamma^2\cos^2\theta\right)^{1/2}$ with $\gamma=H_{C2}^{//}/H_{C2}^{\perp}$ respectively. The inset shows a close-up of the region around 90°. (c) Superconductivity of $TaIrTe_4$ single crystal before and after capping with 10 nm thick ferromagnetic Ni film on surface. The dashed gray line marks the temperature of onset $T_c$.

## METHODS

**STM and STS measurement**. Samples were cleaved in situ at room temperature under the vacuum with pressure better than $1\times10^{-10}$ torr. The cleaved sample was quickly transferred into a Unisoko-1300 STM system for ultralow temperature measurements down to 0.4 K.

**Transport measurement.** The transport measurements are carried out in a PPMS-16 system (Quantum Design), a pulsed high magnetic field equipment at Wuhan National High Magnetic Field Center and anisotropic upper critical field at a dilution refrigerator with vector magnet (Leiden CF450). For electrical transport measurements of the $TaIrTe_4$ samples on (001) plane, the standard four-probe or Hall structure configuration is used. The electric current is always applied parallel to the (001) plane along ***a*** axis in our studies. Two silver paste current electrodes (I+ and I−) are pressed on both ends and across the entire width of the sample, so that the current can homogeneously go through the sample in the length direction [100]. The other two silver paste electrodes are pressed in the middle of crystal as voltage probes. For magnetoresistance (or Hall resistivity) measurements, any additional Hall (or resistive) voltage signals due to the misalignment of the voltage leads have been corrected by reversing the direction of the magnetic field.

## SUPPLEMENTARY DATA

Supplementary data are available at *NSR* online.

## ACKNOWLEDGEMENTS

We acknowledge Xincheng Xie, Ji Feng, Haiwen Liu, Cheung Chan and Yi Liu for helpful discussions. We thank Xiyao Hu for help in XRD measurements, Cong Wang for preparation of TEM samples, Liang Li, Haoran Ji, Jiawei Zhang, Yanan Li, Pu Yang and Yongjie Li for their help in electrical transport measurements.

## FUNDING

This work was supported by the National Key Research and Development Program of China (Grant No. 2018YFA0305604, No. 2017YFA0303302, and 2016YFA0301604), the National Natural Science Foundation of China (Grant No. 11888101, 11774008, 11574095, 11704414, No. 11574008, and No. 11761161003, No.11974430), Beijing Natural Science Foundation(Grant No. Z180010), the Science Foundation of China University of Petroleum, Beijing (Grant No. 2462017YJRC012, 2462018BJC005), and the Strategic Priority Research Program of Chinese Academy of Sciences (Grant No. XDB28000000). Z.Z. was partially supported by the CEM, an NSF MRSEC (Grant DMR-1420451). Work at ORNL was supported by the US Department of Energy, Office of Science, Basic Energy Sciences, Materials Sciences and Engineering Division (J.Y.). D.M. acknowledges support from the Gordon and Betty Moore Foundation's EPiQS Initiative through Grant No. GBMF4416. B.Y. acknowledges the financial support by the Willner Family Leadership Institute for the Weizmann Institute of Science, the Benoziyo Endowment Fund for the Advancement of Science, Ruth and Herman Albert Scholars Program for New Scientists, the European Research Council (ERC) under the European Union's Horizon 2020 research and innovation programme (grant agreement No. 815869).

SUPPLEMENTARY INFORMATION

**Figures**

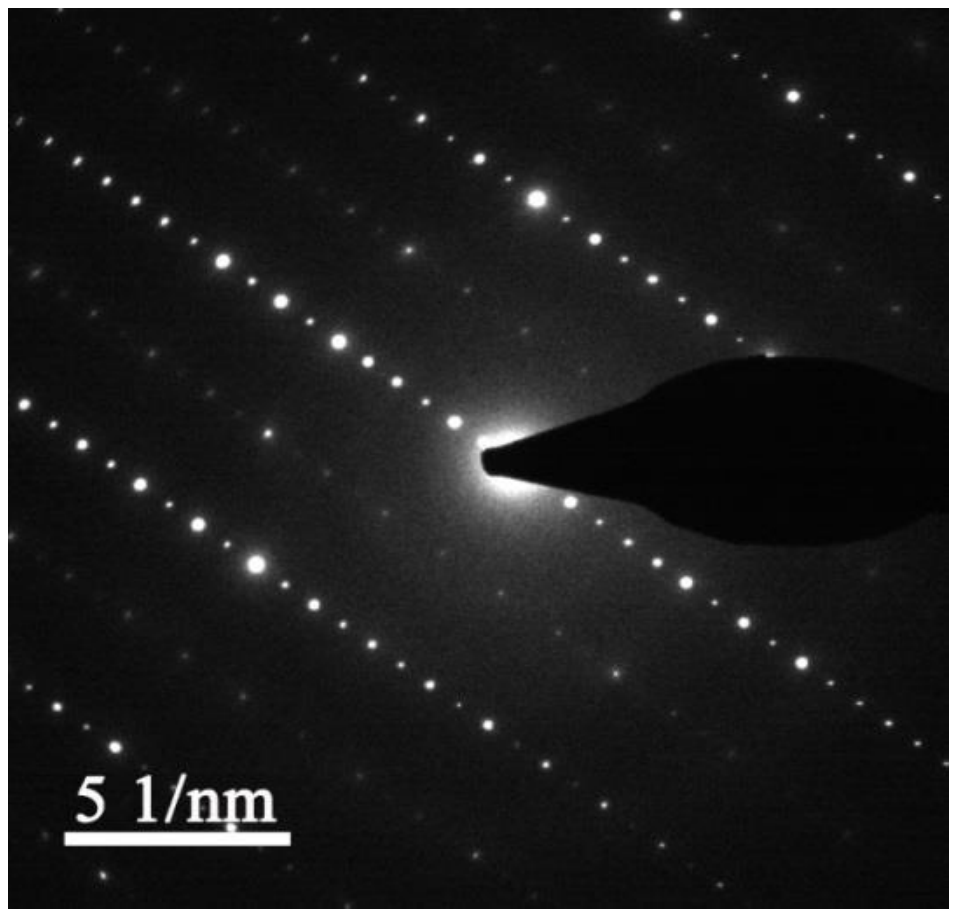

FIG. S1. Electron diffraction image looking down the [100] zone axis showing the reciprocal lattice of $TaIrTe_4$.

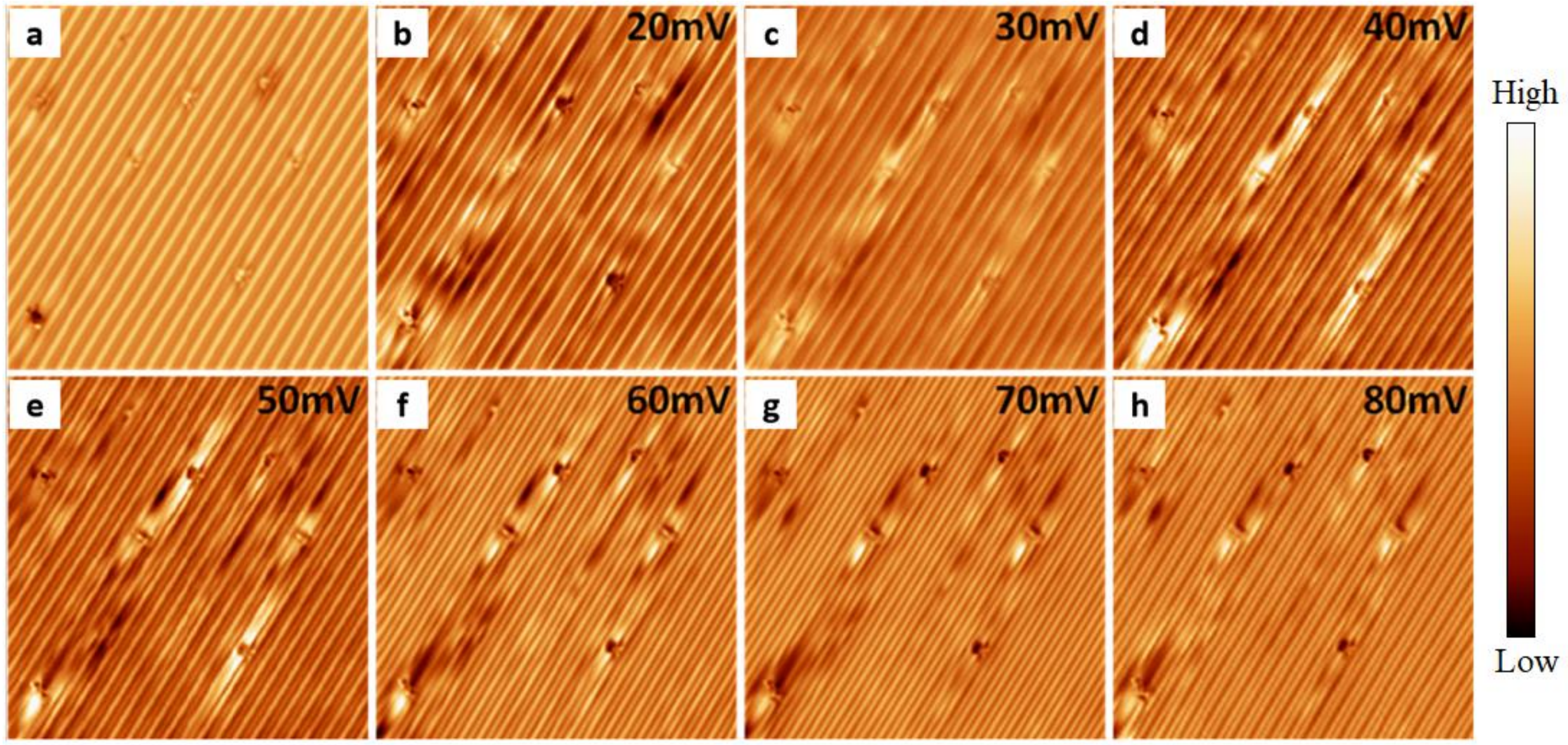

FIG. S2. (a) Topography of a chosen surface 40 nm $\times$ 40 nm and (b)- (h) the corresponding differential conductance mappings at different biases ($I_{set}$= 250 pA, bias modulation amplitude of 2.4 mV).

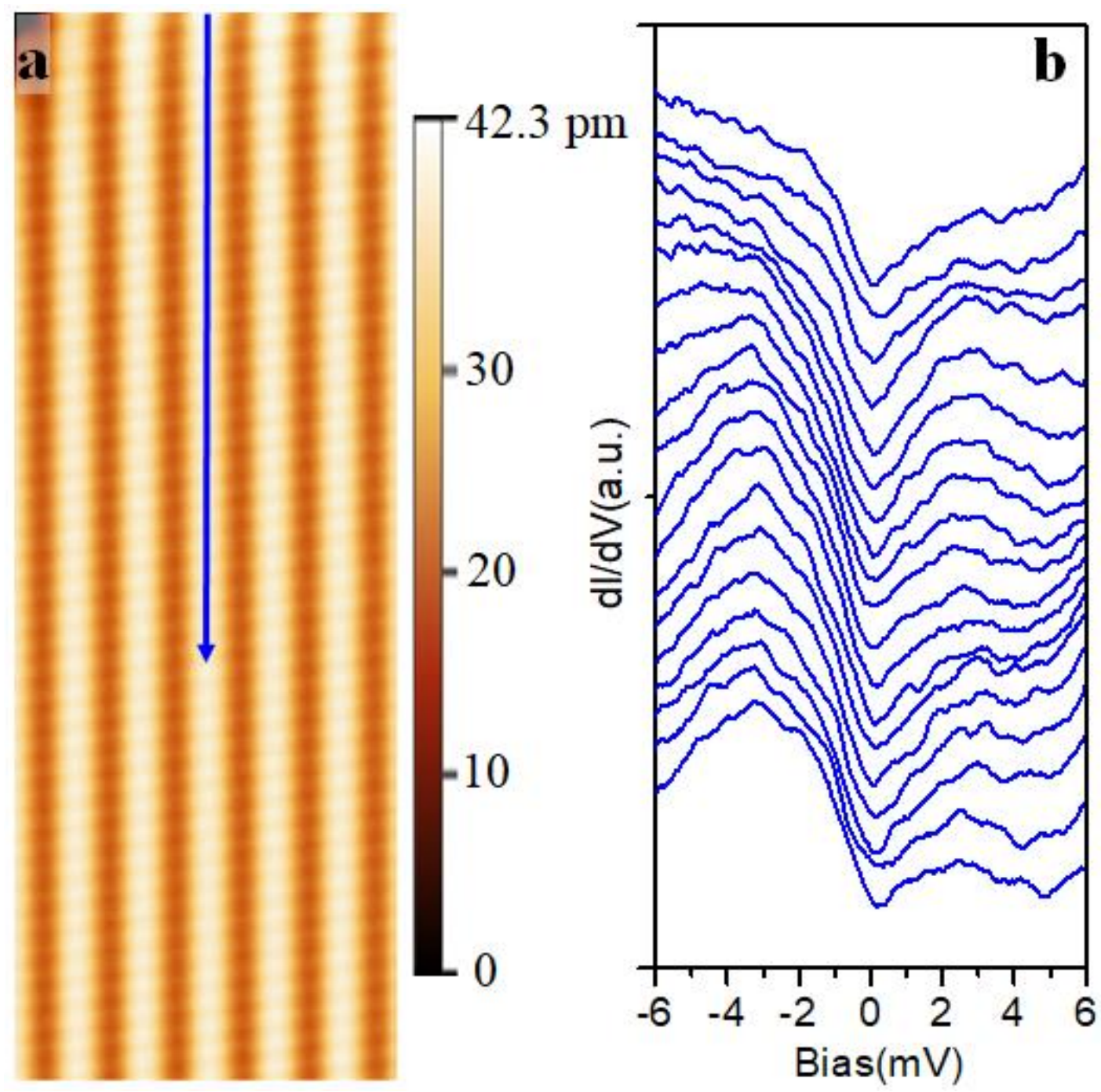

FIG. S3. (a) STM image shows a surface area, image size: 10 nm $\times$ 30 nm. (b) A series of spectroscopic survey taken along the blue line in (a) panel. All d$I$/d$V$ tunneling spectra are measured at 0.4 K with a bias voltage of -10 mV, a tunneling current of 500 pA. The bias modulation is set at 300 μV.

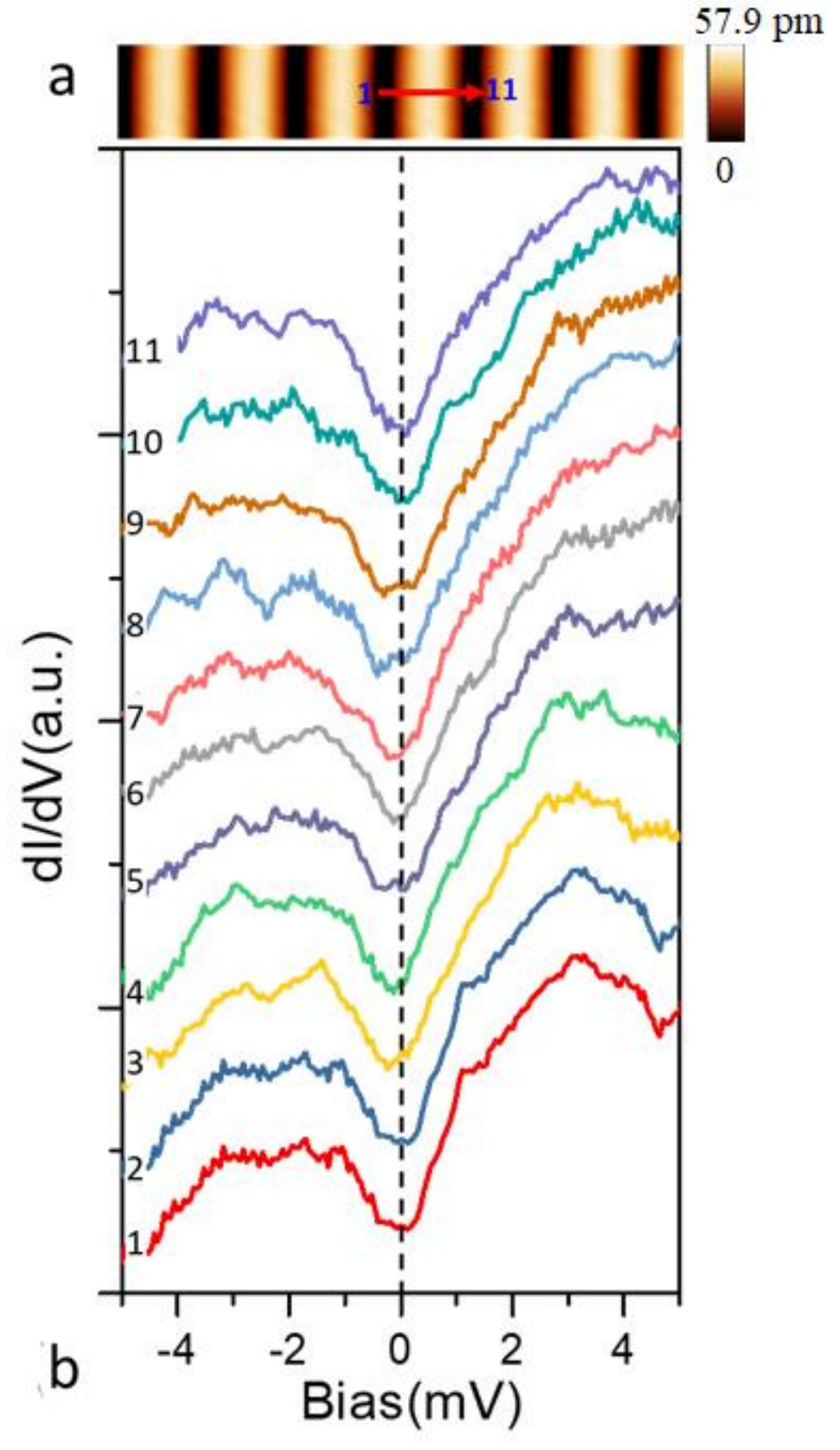

Fig. S4. (a) STM image shows a surface area with periodical 1D chains. (b) A series of spectroscopic survey taken along the red arrow in (a) panel by across a 1D chain.

**Supplementary discussions on exclusion of other possible mechanisms of the STS gap formation**

**1. Exclusion of pseudogap**

In such a narrow temperature range of 0.4-1.1 K, extremely small temperature variation can induce significant change of gap shape and eventually gap feature is vanished at $T_c$, ruling out the possibility of pseudogap which is insensitive to the variation of temperature.

**2. Exclusion of charge density wave (CDW) gap**

By high-resolution STM measurements, CDW is not observed in atomically resolved images at 0.4 K and 4 K, suggesting that the gap should not be a CDW gap.

**3. The STS and transport measurement results together evidence the observation of superconductivity**

The superconductivity of gap feature can be supported by the evolution of tunneling spectra with increasing temperature and external magnetic field (Fig. 3(c)-(d) in main text) respectively, exhibiting a typical behavior of superconducting gap. More convincingly, the critical temperature and critical magnetic field obtained from STS investigation match well with the results of transport measurement as shown in Fig. 4 (main text).

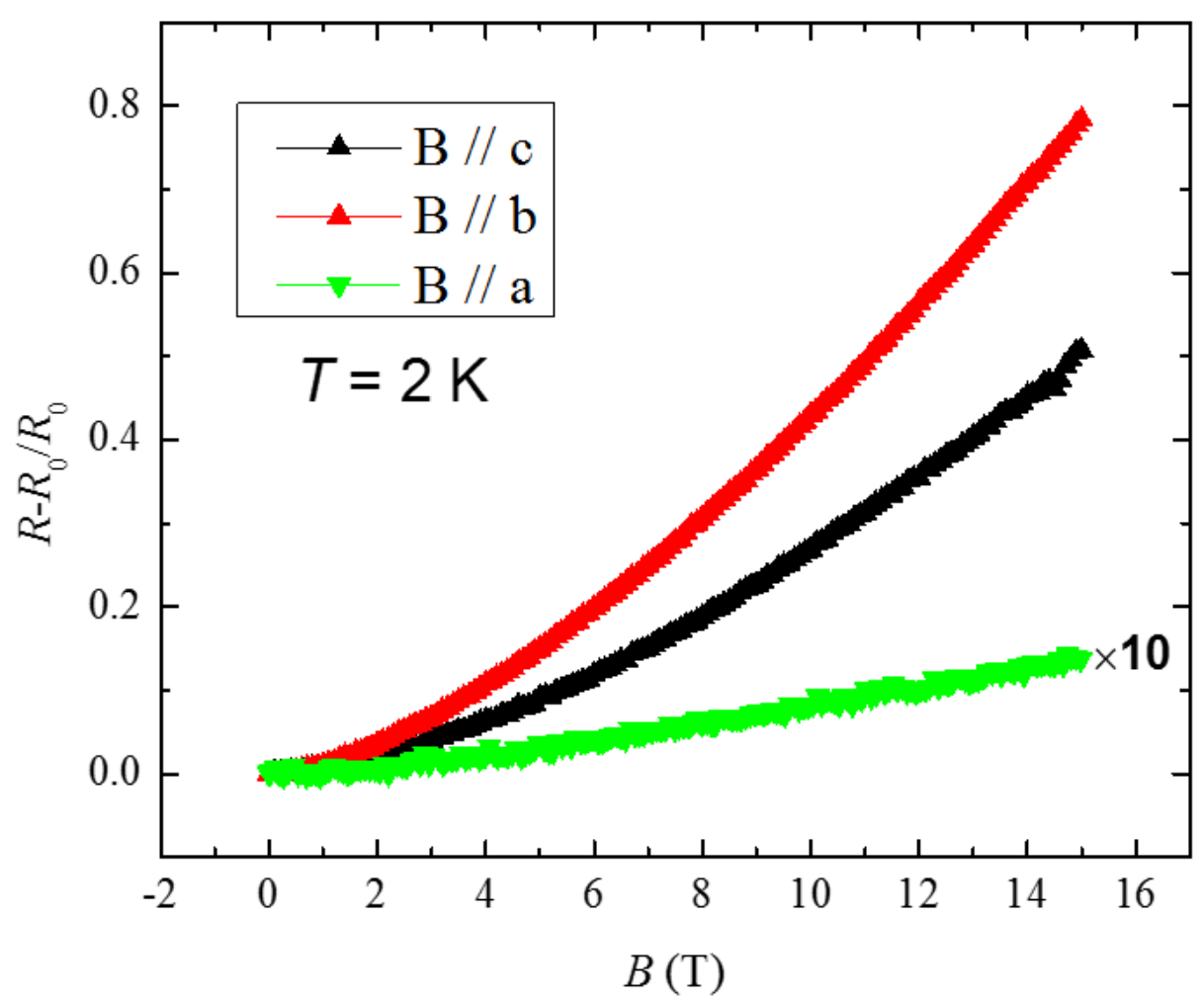

FIG. S5. Magnetoresistance of S1 for three orientations of the magnetic field at 2 K.

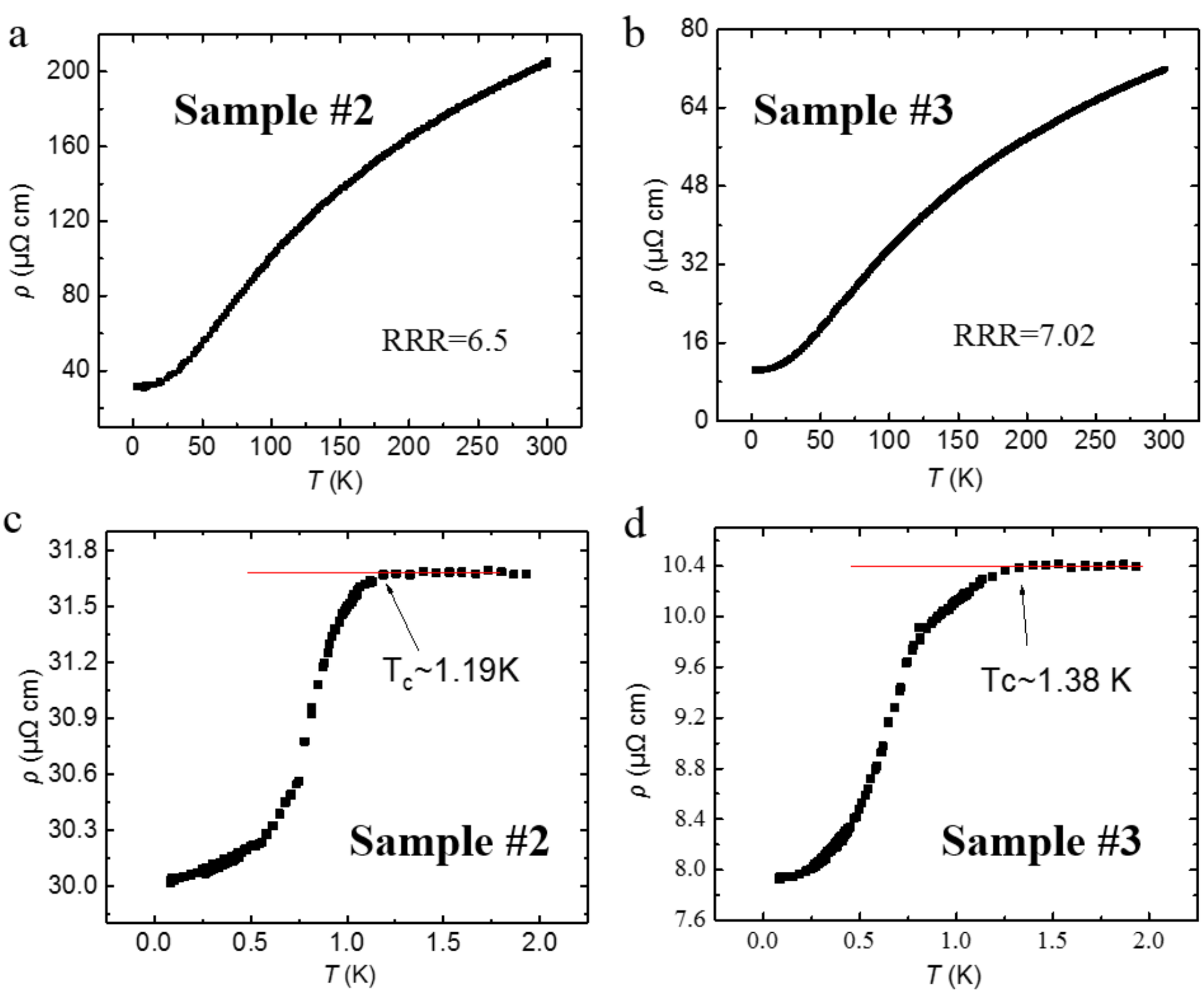

FIG. S6. Superconducting properties of S2, S3. $\rho(T)$ curves of S2(**a**), S3(**b**) from 2 K to 300 K at zero magnetic field. (c)**,** (d) $\rho(T)$ curves below 2 K. The resistivity drops locate at 1.19 K, 1.38 K, respectively.

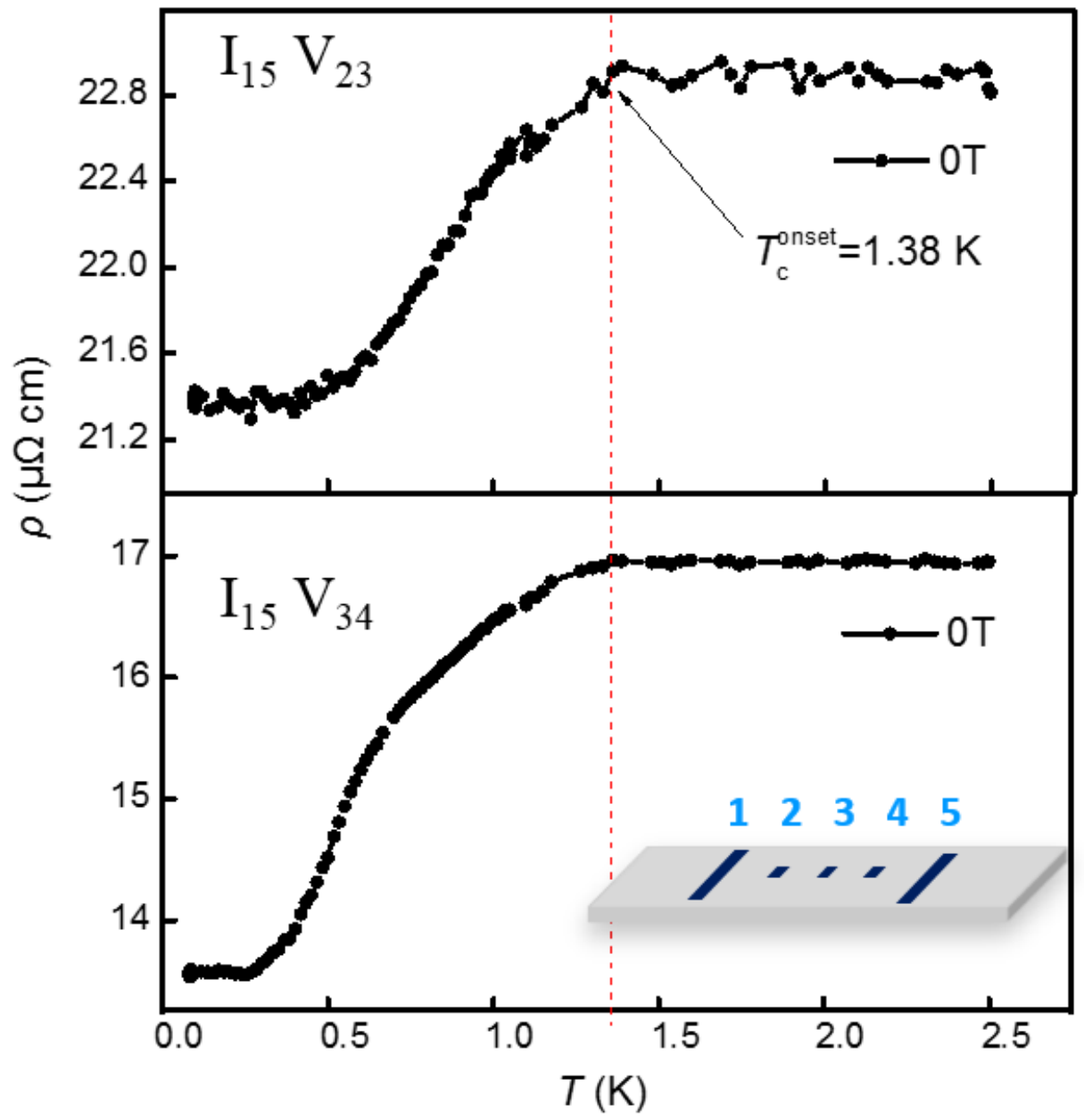

FIG. S7. $\rho(T)$ curves of S3. Five electrodes are made on a long striped $TaIrTe_4$ sample. Standard four-probe method is used to measure the transport property of different regions in one $TaIrTe_4$ sample.

The current sides are 1 and 5. The voltage sides are 2, 3 (upper panel) and 3, 4 (lower panel). $T_c^{onset}$ for both regions is similar and about 1.38 K. This excludes the macroscopic superconducting phase separation in $TaIrTe_4$ samples. The inset shows the measurement structure schematic diagram.

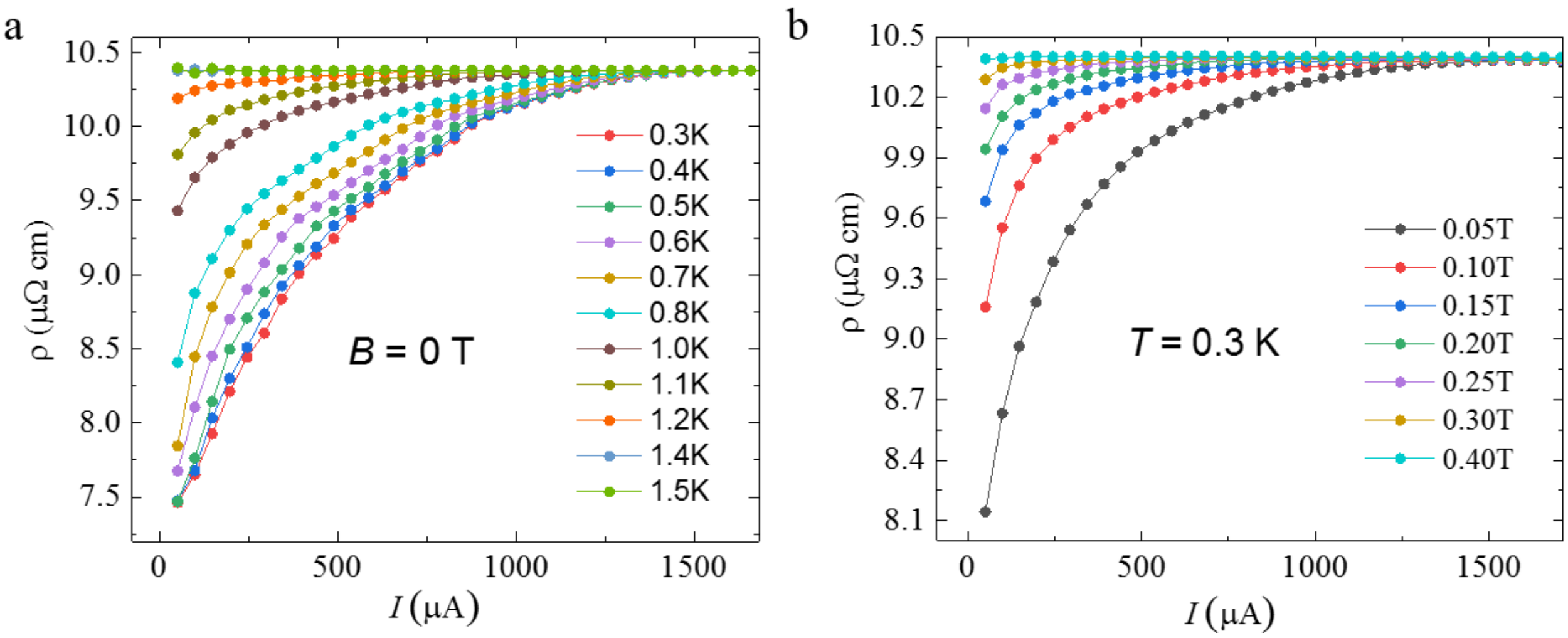

FIG. S8. The sample was obtained by mechanical exfoliation from Sample 3. (a) $R(I)$ curves of 30 μm thick $TaIrTe_4$ measured at temperatures ranging from 0.3 K to 1.5 K at $B = 0$ T. (b) $R(I)$ curves of 30 μm thick $TaIrTe_4$ measured at magnetic field ranging from 0.05 T to 0.4 T for $T = 0.3$ K.

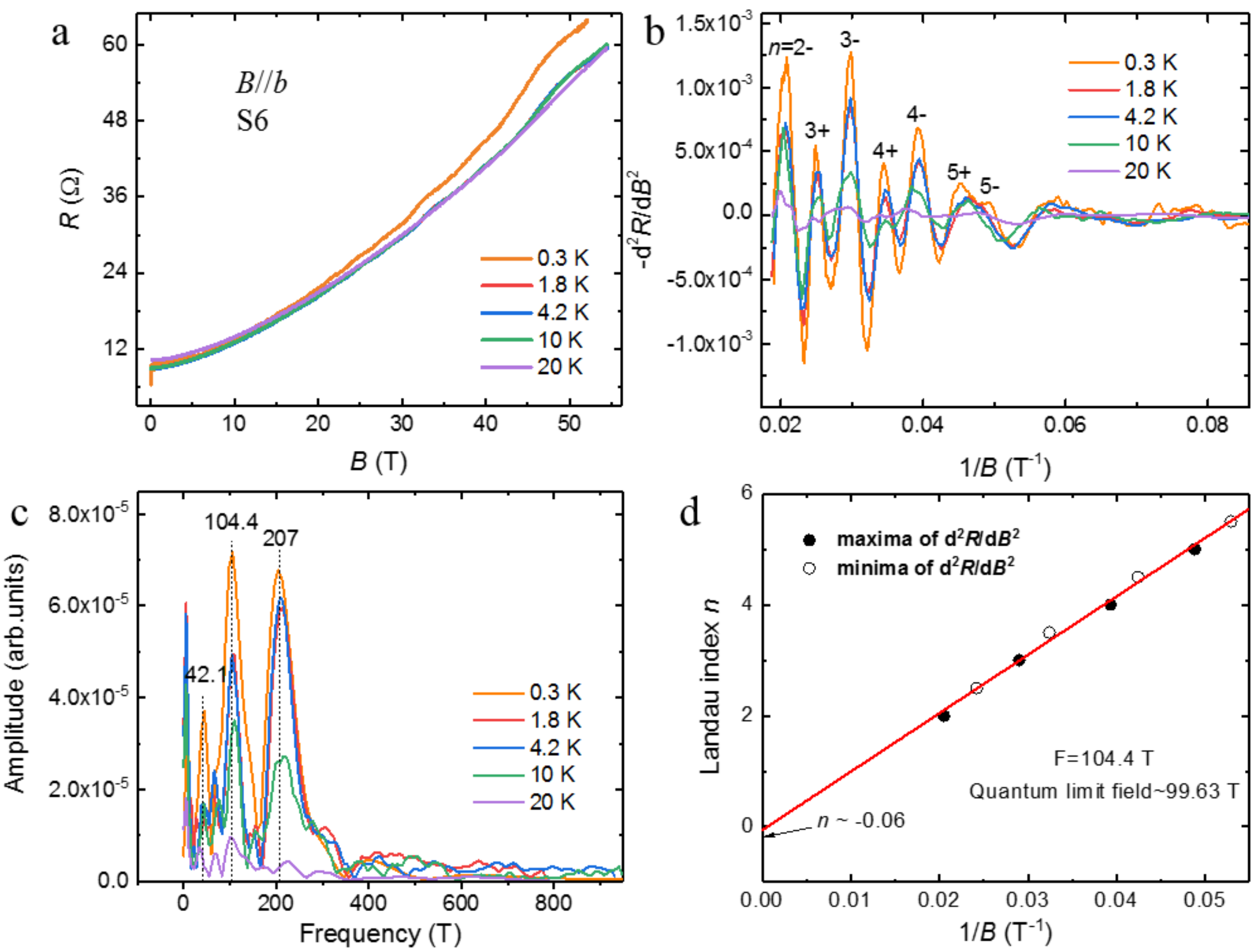

FIG. S9. Quantum oscillations in $TaIrTe_4$ single crystals (S5) with magnetic field perpendicular to the *ac*

plane ($B//b$). (a) Magnetic field (up to 54.5 T) dependence of resistivity at different temperatures. (b) Second derivative of the $\rho(B)$ data in (a) vs. $1/B$. (c) FFT analysis with two major frequencies (42.1 T and 104.4 T) for $d^2R/dB^2$ vs. $1/B$ in (b). (d) Landau level indices ($n$) as a function of $1/B$.

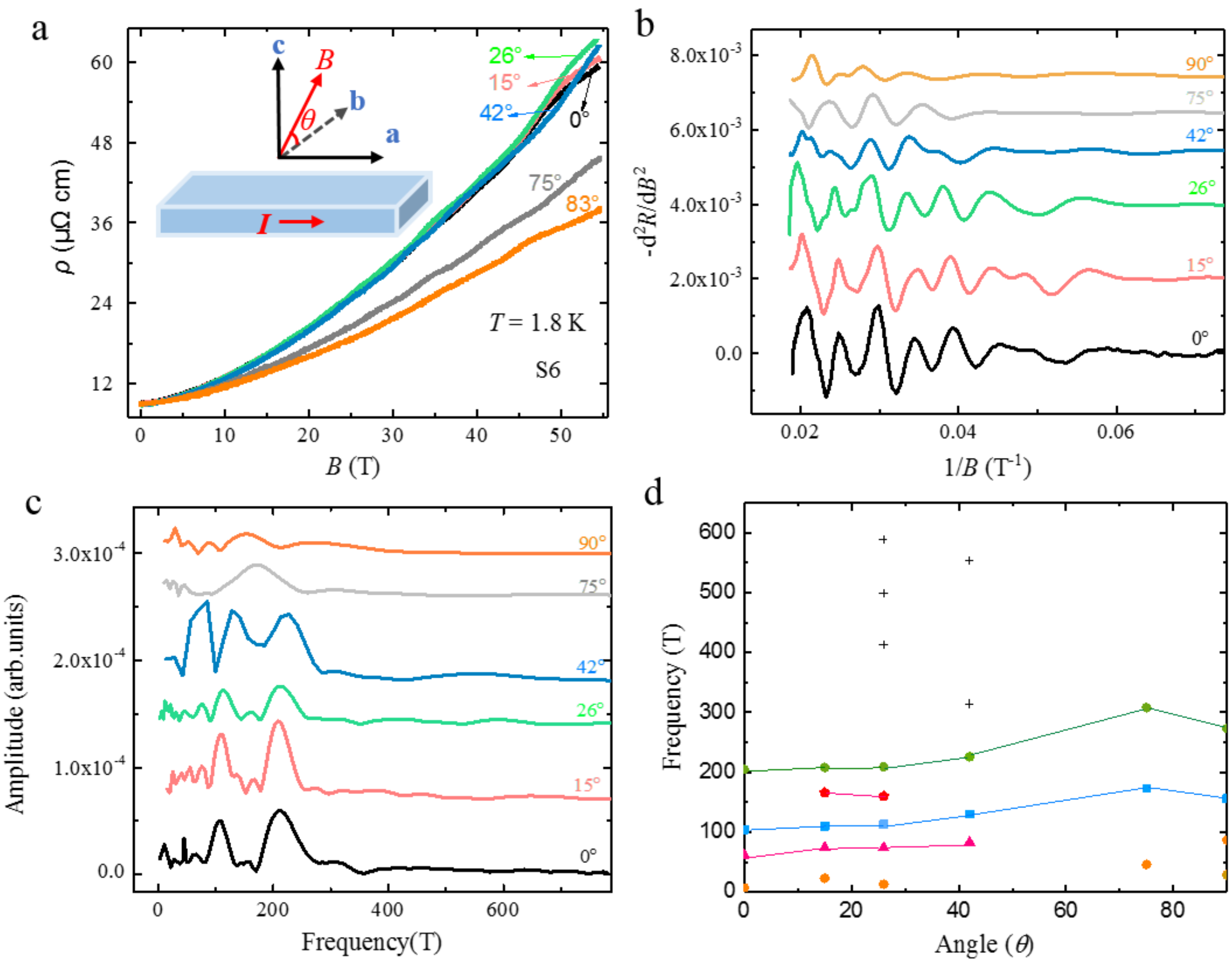

FIG. S10. Angular dependence of SdH oscillations of S5 by rotating the sample from [010] direction to [001] direction at different magnetic field directions. (a) $\rho(B)$ curves at different magnetic field angles from $B_{[010]}$ ($\theta = 0°$, $B$ is parallel to the $b$ axis) to $B_{[001]}$ ($\theta = 90°$, $B$ is parallel to the $c$ axis). (b) Second derivative of the $R(B)$ data in (a) vs. $1/B$. (c) FFT analysis for $d^2R/dB^2$ vs. $1/B$ in (b). (d) Angular dependence of the SdH frequencies determined in transport data.

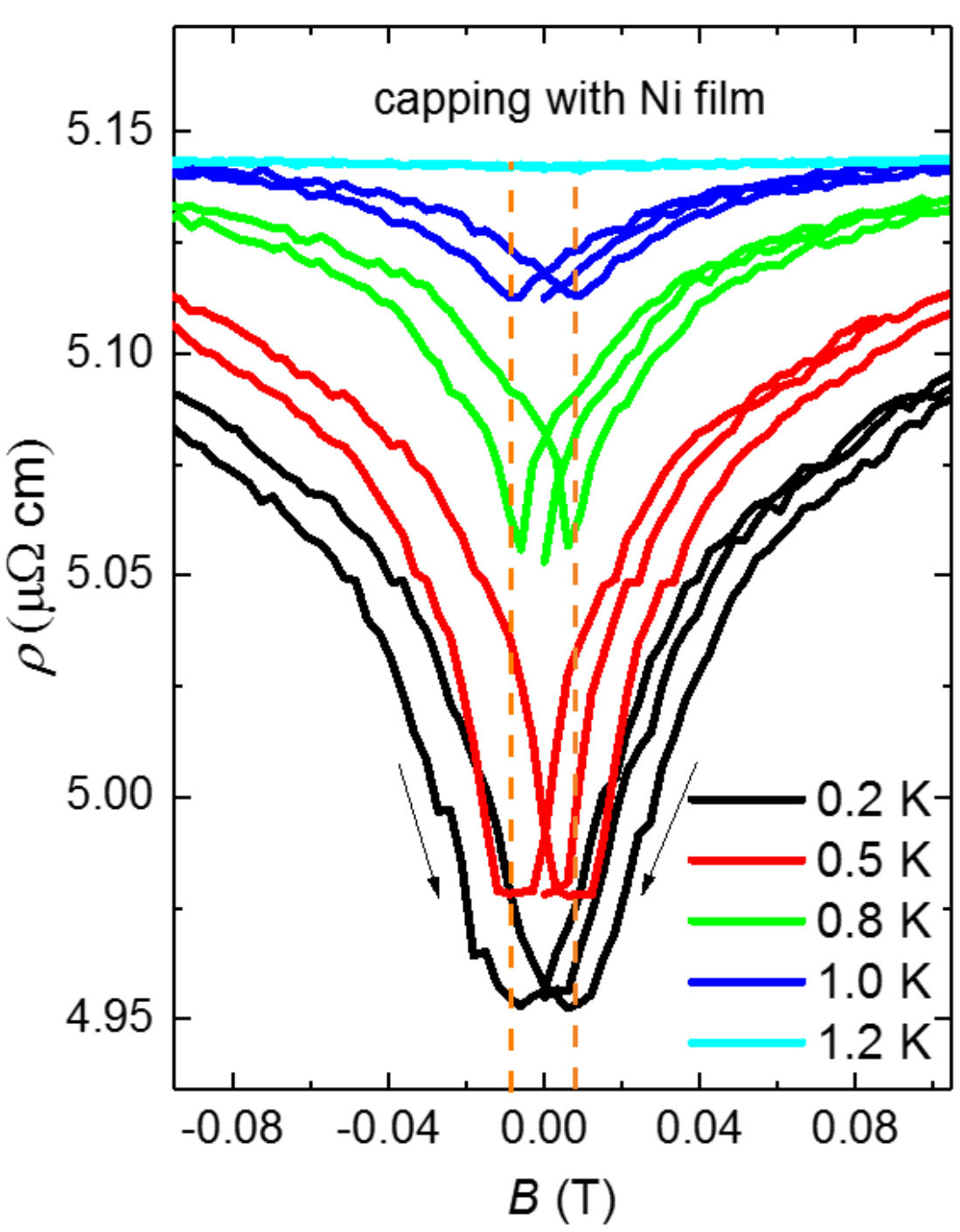

FIG. S11. Magneto-resistance of $TaIrTe_4$ capping with Ni film from 0.2 K to 1.2 K reveals the hysteretic behavior in different sweep directions. The black arrows indicate the magnetic field sweep direction.

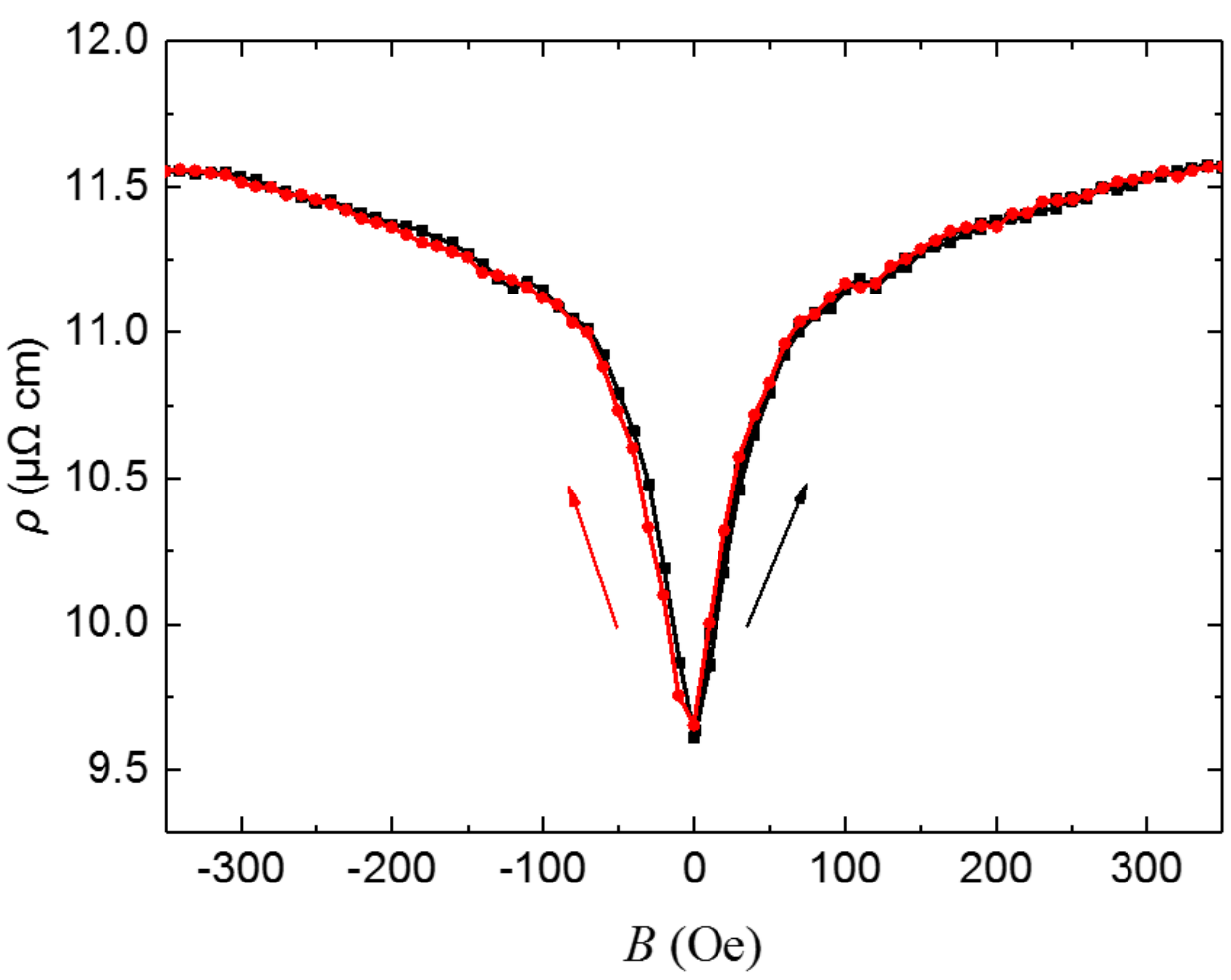

FIG. S12. Magnetoresistance of S6 at 0.5 K. No oscillation and hysteretic behavior was found from different sweep directions at low field region. This excludes the possibility of magnetic phase transitions in $TaIrTe_4$ at low temperature.

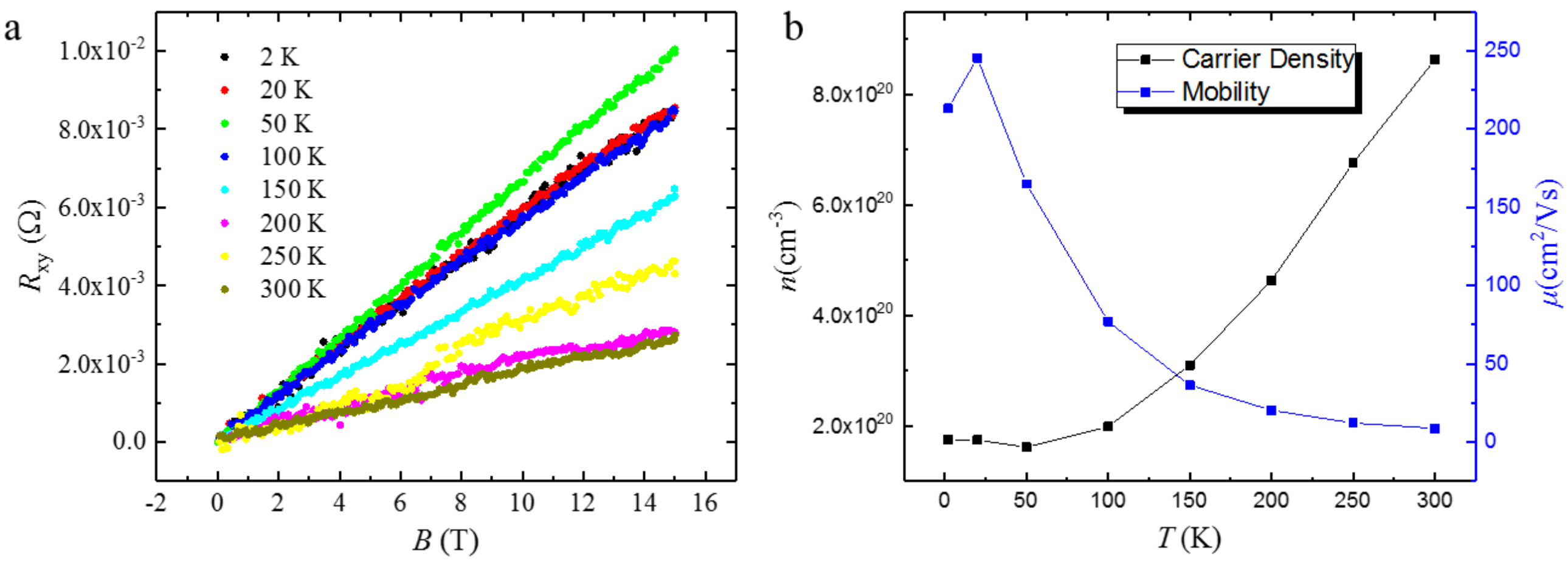

FIG. S13. Hall results of S1. (a) Hall resistance ($R_{xy}$) varies with magnetic field at different temperatures from 2 K to 300 K. The field is applied perpendicular to the *ab* plane (*B*//*c*) and the current is along the *a*-axis of the orthorhombic crystal structure (*I*//*a*). Any additional resistance due to the misalignment of the voltage leads or thermal effect has been removed by averaging the $R_{xx}$ data over positive and negative field directions. (b) Calculated carrier density ($n$) and mobility ($\mu$) at different temperatures.

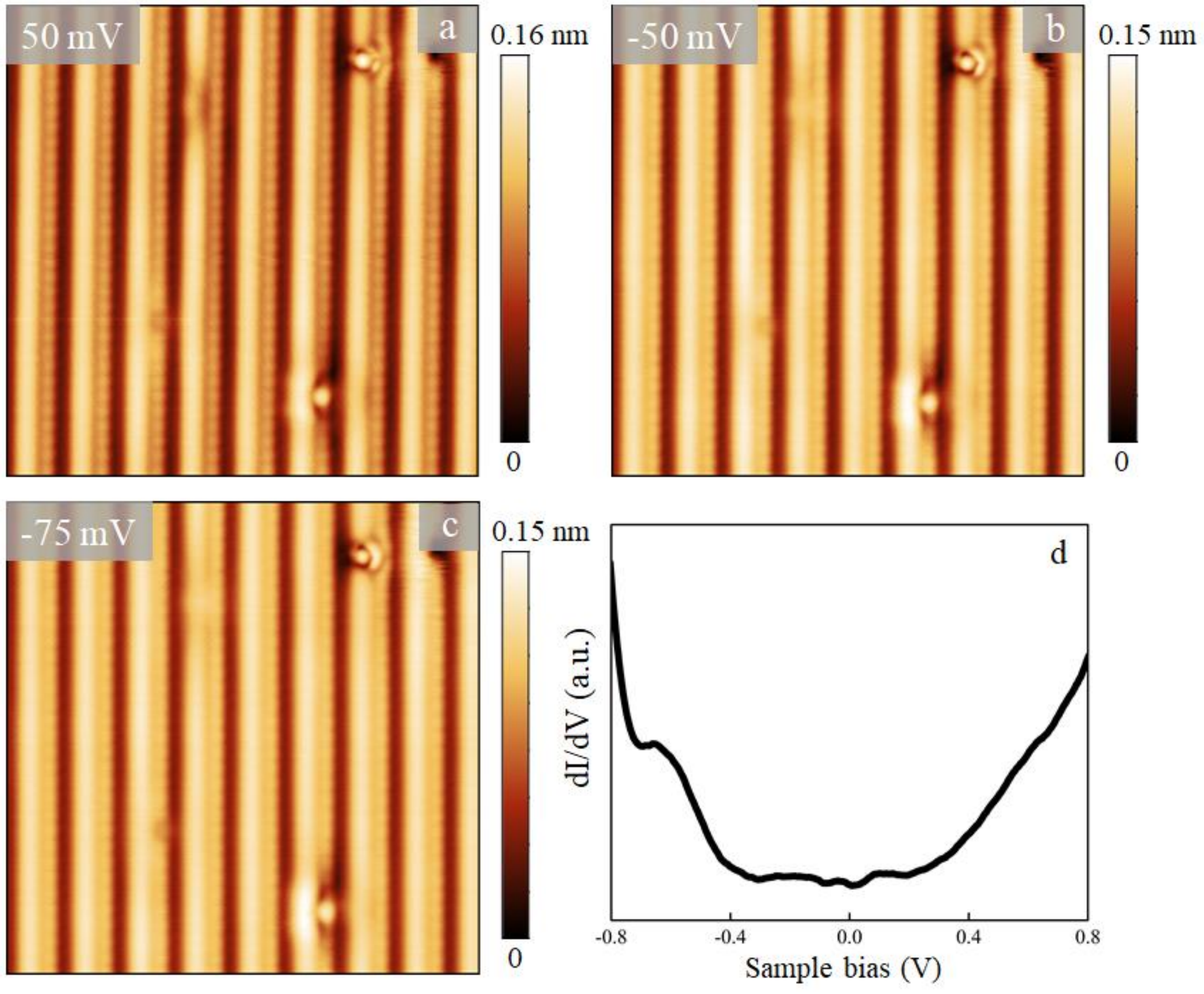

Fig. S14. (a)-(c) The atomically resolved images acquired at different bias. Size: $10\times10$ nm$^2$, $I_t$=300 pA.

(d) Large-energy-scale dI/dV spectrum. The bias modulation and tunneling current are set at 8 mV and 200 pA respectively.

The experimental evidences on surface superconductivity in $TaIrTe_4$ are summarized as **Table I**.

| **Conclusion** | **Experimental evidence** | **Note** |
|---|---|---|
| Superconductivity | Superconducting gap which can be suppressed by increasing magnetic field and temperature (Fig.3) | Evidence of superconductivity |
| | Superconducting resistance drops which can be suppressed by increasing magnetic field, temperature and current (Fig.4b, Fig.4c, Fig.5a, Fig. S8) | Evidence of superconductivity |
| Homogeneity of superconductivity | XRD(Fig.1a), STEM(Fig.1b) , STM(Fig.1d,1e,Fig.3a) | Evidence of high quality single crystal |
| | Multi-electrodes transport results (Fig.S7) | Exclude macroscopic phase separation |
| | Uniform superconducting gap in whole surface(Fig.S3) | Exclude minority phase region |
| The superconductivity comes from the surface state. | Nearly thickness-independent critical current. (Fig. 5a) | Fully exclude the possibility of bulk superconductivity |
| | The angular dependence of the upper critical field (Fig.5b) | A cusp-like peak is qualitatively distinct from the 3D model |
| | The superconducting gap is detected by surface sensitive STM (Fig.3, Fig.S3 and Fig.S4) | The sample surface is superconducting. |
| The possibility of nontrivial topological superconductivity | Fermi arc states observed by STS (Fig.2) | Topological non-trivial surface state |
| | Ferromagnetic Ni film has little effect on the onset $T_c$ of $TaIrTe_4$ (Fig.5c,Fig.S11) | Suggesting the possibility of topological superconducting paring symmetry in $TaIrTe_4$ |
| | $h^*(T/T_c)$ relation is close to that of a polar *p*-wave state (Fig. 4c inset) | Suggesting the possibility of topological superconducting paring symmetry in $TaIrTe_4$ |
| | Superconducting gap becomes smaller and shallower by approaching to the broken end (Fig. 3f) | Suggesting that the pairing order might be unconventional |
| Quasi 1D feature | STEM (Fig.1b) , STM (Fig.1d, Fig.1e, Fig.3a) | Quasi-1D structure |
| | Fermi arc states observed by STS (Fig.2) | Quasi-1D behavior on the surface |
| | The anisotropy of upper critical field(Fig.4f) and magnetoresistance (Fig.S5) | Quasi-1D superconductivity and magnetoresistance characteristic |